\documentclass{ismdproc}

\begin{document}
\title{Measuring the thermalization time}

\author{{\slshape Piotr Bo\.zek$^{1,2}$\footnote{speaker, email: piotr.bozek@ifj.edu.pl}\ , Iwona Wyskiel-Piekarska$^{1}$}\\[1ex]
$^1$The H. Niewodnicza\'nski Institute of Nuclear Physics,
PL-31342 Krak\'ow, Poland\\
$^2$Institute of Physics, Rzesz\'ow University, 
PL-35959 Rzesz\'ow, Poland }

\contribID{xy}  
\confID{yz}
\acronym{ISMD2010}
\doi            

\maketitle

\begin{abstract}
 A new method of measuring the pressure anisotropy in the early stage of 
heavy-ion collisions is discussed. It is shown that the transverse momentum 
spectra, elliptic flow and interferometry radii are not sensitive to the early anisotropy. We
propose the directed flow as a measure of the early transverse and longitudinal 
pressures. Calculations indicate that the transverse and 
longitudinal pressures equilibrate in less than $0.25$~fm/c.
\end{abstract}

\section{Introduction}

The dynamics of ultrarelativistic heavy-ion collisions can be described using 
the hydrodynamic model \cite{Kolb:2003dz,Hirano:2008aj,Ollitrault:2010tn}.
 The fireball created in the collision
consists of a dense matter close to equilibrium. The fireball expands 
hydrodynamically and a collective velocity field builds up. Particle production
 happens through the statistical emission on the freeze-out hypersurface. The
 spectra of emitted particles reflect the underlying collective velocity
 field of the fluid. In particular,  the
transverse momentum spectra and the elliptic flow coefficient
 for different particles species can be described quantitatively 
using hydrodynamics.

The hydrodynamic model requires that the fluid is in local thermal equilibrium.
Limitations of this ideal fluid picture are discussed in terms of possible
 viscous corrections to the hydrodynamic equations or in terms of the 
limited time interval 
 when the collective expansion takes place. Quantitative estimates show 
that the shear viscosity coefficient is  small 
\cite{Romatschke:2009im,Teaney:2009qa} and the
viscosity
corrections influence significantly only the elliptic flow results.
The length of the hydrodynamic stage is determined by the initial time $\tau_0$
needed to form the fireball.  Calculations using a hard equation of state 
of the hot matter do not require a short initial time to describe the spectra,
but to describe the femtoscopy results a short formation time 
$\tau_0=0.1$-$0.25$~fm/c is preferred \cite{Broniowski:2008vp,Pratt:2008qv}.
 In a boost
 invariant geometry, 
observables such as the transverse momentum spectra, elliptic flow and 
interferometry radii reflect the accumulated transverse flow of the fluid.
The longitudinal expansion sets the cooling rate of the fluid. However,  
during a short expansion, scenarios with different longitudinal pressures,
 and hence, different cooling rates, lead to  similar transverse
 collective flows
\cite{Broniowski:2008qk,Vredevoogd:2008id}.

The initial time $\tau_0$, when the transverse expansion starts is not
equivalent to the time required for the equilibration of the system.
A necessary condition for the thermalization is that the transverse and
 longitudinal pressures in the fluid become similar. One expects
 that in the initial stage of the collision,
the longitudinal pressure is smaller than the transverse one. Possible 
implementations of such a transient, nonequilibrium pressure anisotropy
in hydrodynamic equations have been discussed 
\cite{Bozek:2007di,Florkowski:2010cf,Martinez:2010sc}.  However, 
as noted above, the 
presence of the pressure anisotropy in the initial stage does not change 
the transverse spectra, elliptic flow or femtoscopy results. It has been
 proposed to look  at photon or dilepton emission 
 instead \cite{Mauricio:2007vz,Schenke:2006yp,Dusling:2008xj}, but
drawing conclusions on the thermalization rates 
is difficult due to unknown backgrounds.
We show that the directed flow of particles is an observable extremely 
sensitive to the initial pressure anisotropy \cite{Bozek:2010aj} 
and it can be used to estimate the thermalization time.

\section{Calculations}

The spectra of emitted particles are written as
\begin{equation}
\frac{dN}{d^2p_\perp dy}=\frac{dN}{2\pi p_\perp dp_\perp dy}\left( 1+2 v_1 \cos(\phi)+2 v_2 \cos(2\phi)+\dots \right) ,
\end{equation}
\begin{figure}[hb]
{\includegraphics[width=0.52\textwidth]{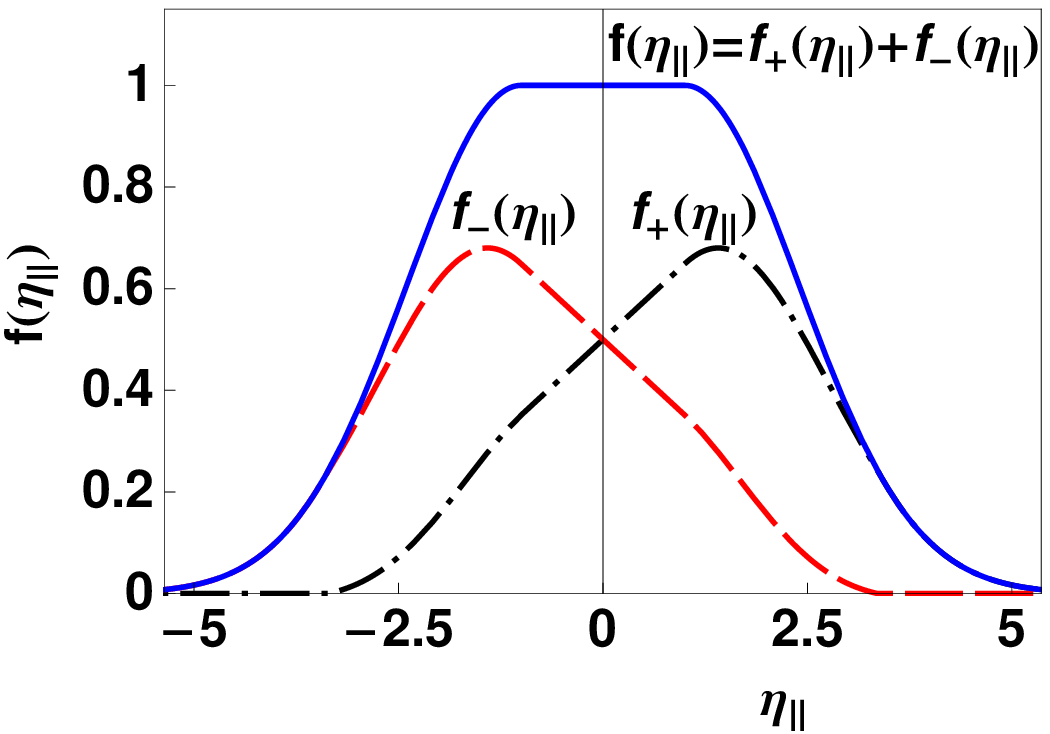}} 
\hfill
{\includegraphics[width=0.45\textwidth]{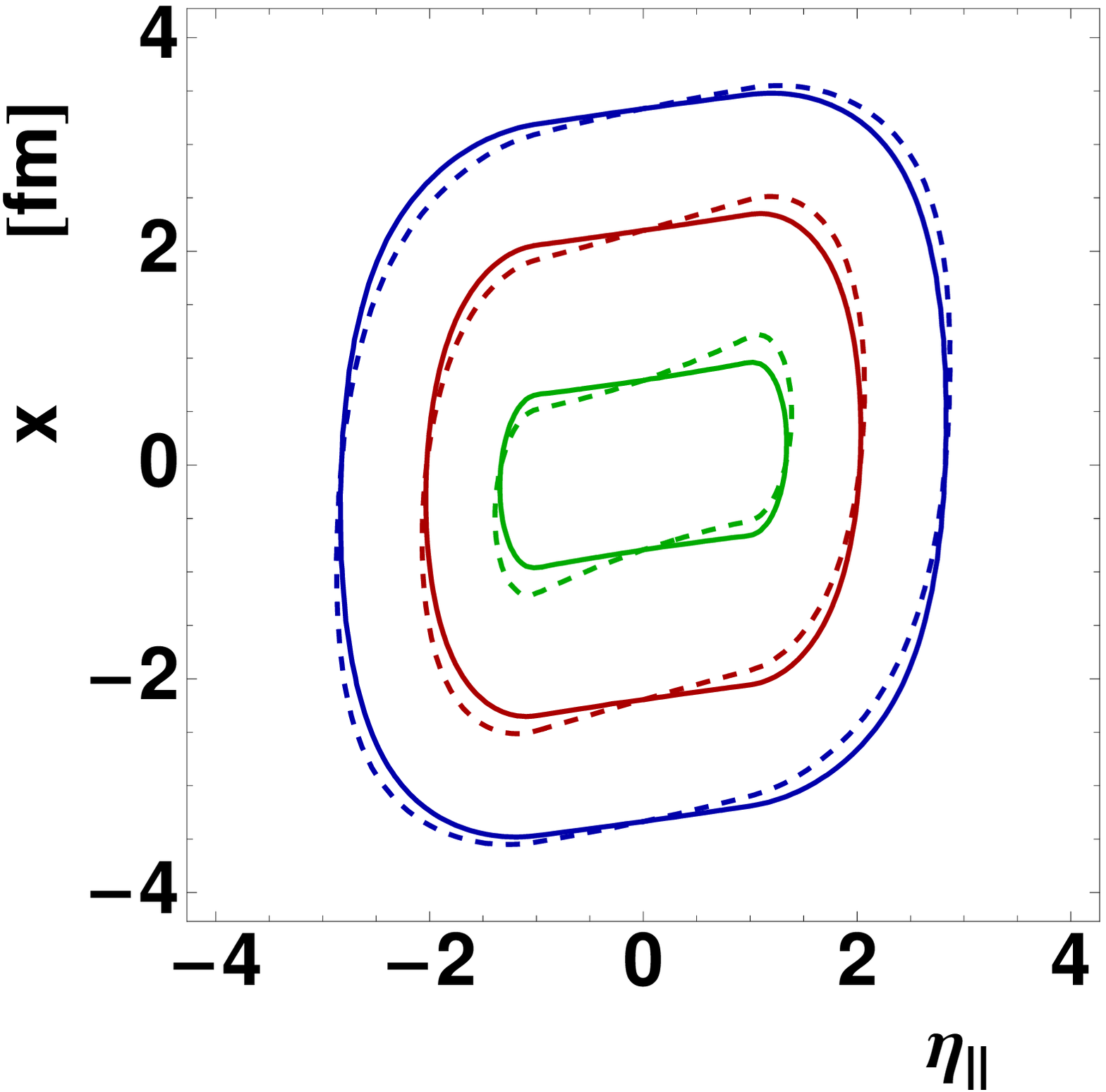}}
\caption{(left panel) The initial energy density distribution 
in space-time rapidity. The energy density is a sum of a two 
components from forward and backward going wounded nucleons \cite{Bozek:2010bi}.
 (right  panel) Contour plot of the initial energy density 
in the ($x$-$\eta_\parallel$) plane.
 The tilt of the source away from the collisions axis results from the
 forward (backward) peaked emission from wounded nucleons. The solid 
 and dashed lines represent the two extreme estimates of the source tilt 
in the Glauber model.}
\label{fig:ini}
\end{figure}
where $v_1$ and $v_2$ are the directed and elliptic flow coefficients.
In symmetric collisions the directed flow coefficient $v_1$ is zero at central
 rapidities, and becomes negative (positive) at forward (backward) rapidities
\cite{Abelev:2008jga}. The formation of directed flow in ultrarelativistic 
collisions requires a mechanism that breaks the symmetry with respect to the 
collisions axis and some
 effective transverse and longitudinal acceleration of the
 fluid elements. The asymmetric emission (Fig. \ref{fig:ini}, left panel) 
in space-time rapidity from the 
forward and backward going wounded nucleons \cite{Bialas:2004su,Adil:2005qn}
results in a tilt of the initial fireball away from the collision axis.
In the right panel of Fig. \ref{fig:ini} is shown the energy density 
for the tilted source. The solid and dashed lines represent two different
source densities calculated in the Glauber model. The difference between 
the two distributions is a measure of the uncertainty in the initial conditions.
 We use the $3+1$ dimensional hydrodynamic model with a parameterization of the equation of state based on lattice QCD results 
\cite{Chojnacki:2007jc}. The model  describes the $p_\perp$ particle 
spectra and the 
pseudorapidity distributions for Au-Au collisions at $\sqrt{s}=200$~GeV
\cite{Bozek:2009ty}. 
In the first fm/c of the hydrodynamic evolution of the tilted source, the
 simultaneous action of the transverse and longitudinal
 pressures generates a negative elliptic flow, similar as 
observed experimentally \cite{Bozek:2010bi}.

\begin{figure}[thb]
{\includegraphics[width=0.46\textwidth]{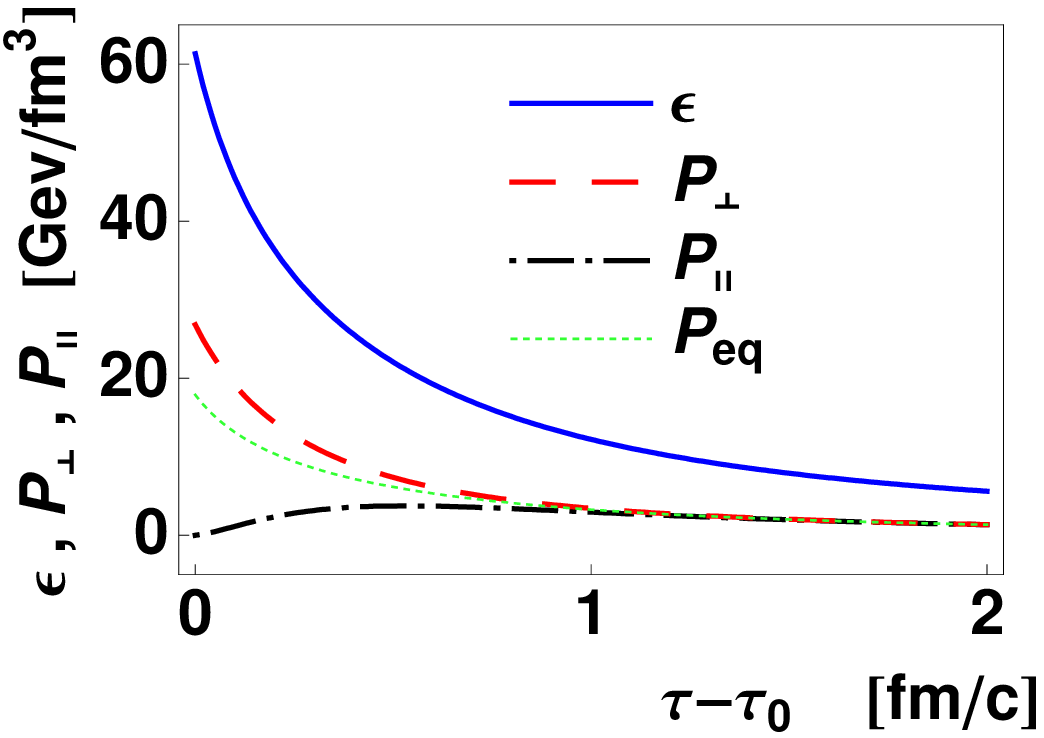}}
{\includegraphics[width=0.54\textwidth]{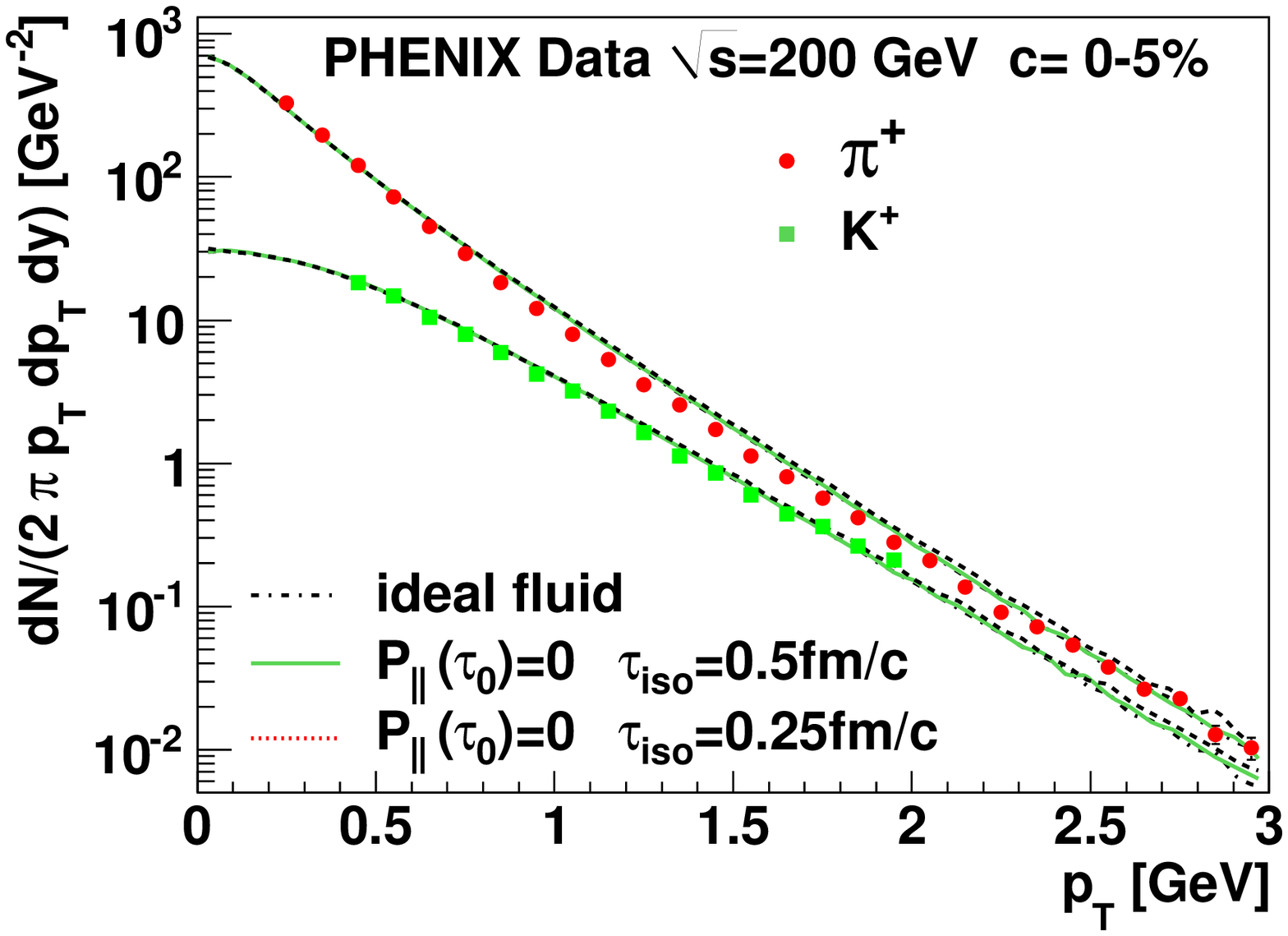}}
\caption{(left panel) Time evolution of the energy density, the 
longitudinal and transverse pressures at the center of the fireball, $P_L(\tau_0)=0$, 
$\tau_{iso}=0.25$~fm/c.
 (right  panel) Pion and kaon transverse momentum spectra calculated in the
 ideal fluid and anisotropic pressures scenarios compared to PHENIX  Collaboration 
data \cite{Adler:2003cb}.}
\label{fig:pt}
\end{figure}

\begin{figure}[htb]
{\includegraphics[width=0.46\textwidth]{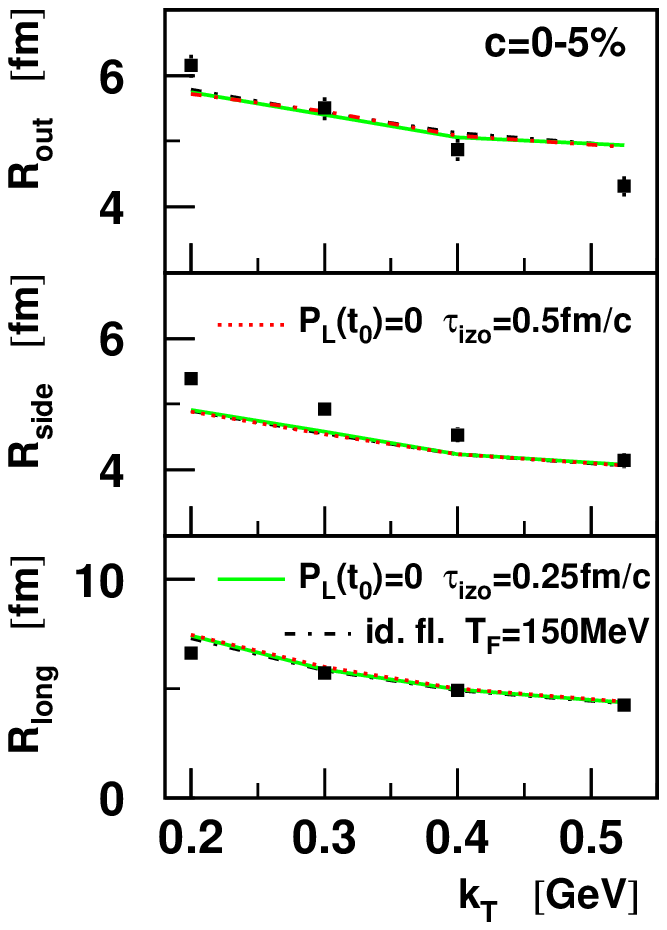}}
{\includegraphics[width=0.54\textwidth]{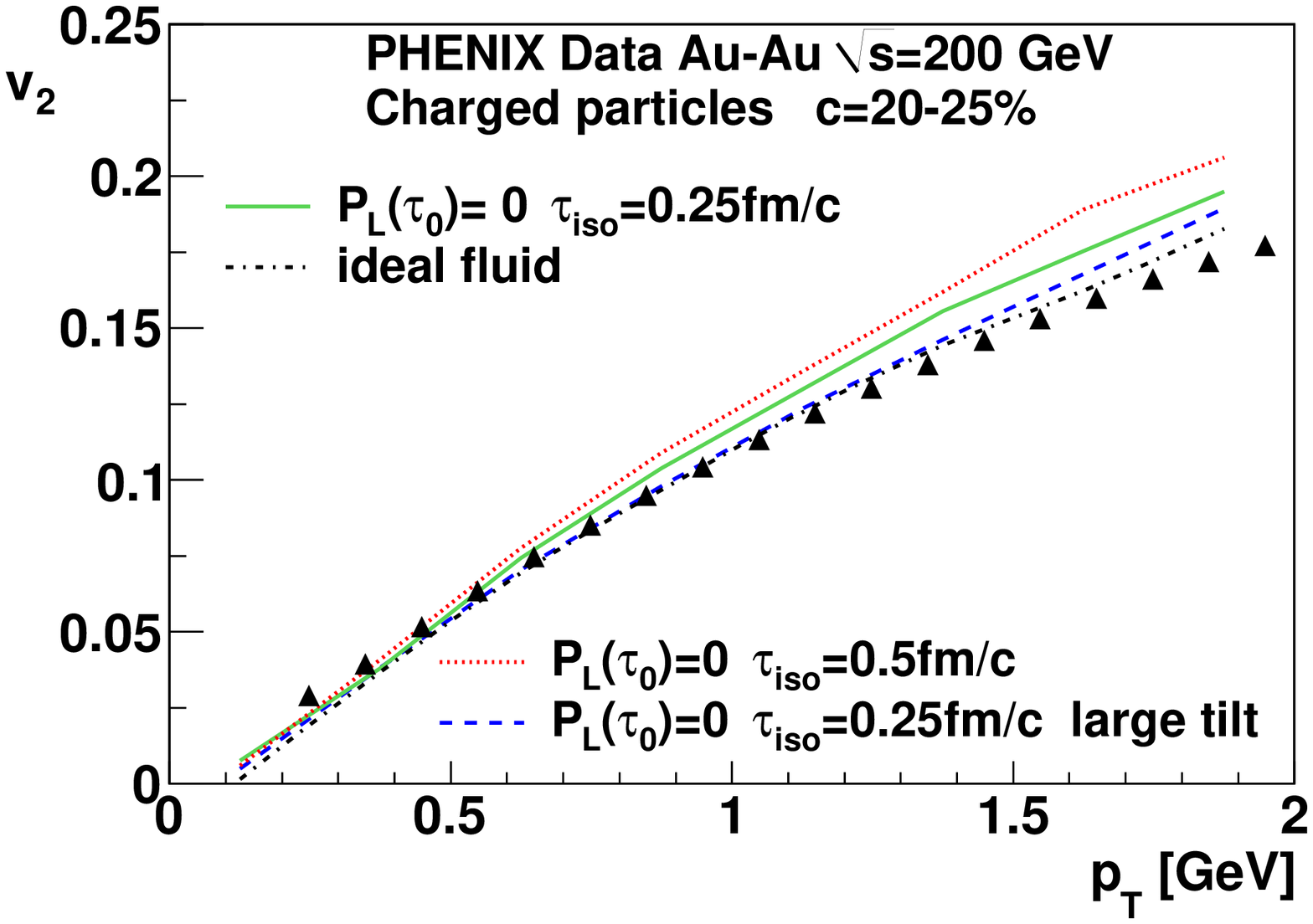}}
\caption{(left panel) Interferometry radii in central Au-Au collisions at 
$\sqrt{s}=200$~GeV calculated for different initial anisotropy scenarios compared to STAR Collaboration data \cite{Adams:2004yc}. (right  panel) Elliptic flow for charged particles as function of transverse momentum for different initial anisotropy scenarios compared to PHENIX Collaboration data \cite{Adare:2010ux}.}
\label{fig:v2}
\end{figure}

We study the dynamics of the system with anisotropic pressures.
In the early stage of the collisions, the energy momentum tensor in
 the local rest frame of the fluid is
\begin{equation}
T^{\mu\nu}=\left( \begin{array}{cccc} \epsilon & 0 & 0 &0\\
0& P_{eq}+\pi/2 & 0 & 0\\
0& 0 & P_{eq}+\pi/2 & 0 \\
0 & 0 & 0 & P_{eq}-\pi 
 \end{array}\right) \ . 
\label{eq:tmunu}
\end{equation}
The correction $\pi$ changes changes the transverse $P_\perp=P_{eq}+\pi/2$ and
the longitudinal $P_\parallel=P_{eq}-\pi$
 pressures. In principle, 
the stress correction to the pressures could be 
calculated from some underlying theory.  The  mechanisms
 of the pressure equilibration is still the subject of studies 
 \cite{Mrowczynski:2005ki,Rebhan:2008uj,Bjoraker:2000cf,Xu:2004mz,Chesler:2009cy,Dusling:2010rm}. Moreover, the models are studied in simplified geometries. 
In the first moments of the collisions
 the
viscous hydrodynamics cannot be used to determine the stress correction, because
 the system is far from equilibrium.
We are interested in the possibility of observing the occurrence of the 
pressure asymmetry, irrespective of its origin. We assume a simple time 
dependence of the correction \cite{Bozek:2007di}
\begin{equation}
\pi(\tau)=P_{eq}(\tau_0)
 e^{(\tau_0-\tau)/\tau_{iso}}\ .
\label{eq:anip}
\end{equation}
It means that initially the longitudinal pressures is zero and that
 the pressure anisotropy decreases with a relaxation time $\tau_{iso}$ 
(Fig. \ref{fig:pt}). The initial density is rescaled to take into account the 
entropy production in the dissipative, anisotropic stage.

We simulate Au-Au collisions at $\sqrt{s}=200$~GeV using the perfect fluid 
hydrodynamics ($\tau_{iso}=0$) and using a  hydrodynamic model with the 
pressure anisotropy  with
 $\tau_{iso}=0.25$ and $0.5$~fm/c. The transverse momentum spectra are
very similar (Fig. \ref{fig:pt}). Also the interferometry radii 
and the elliptic flow coefficient come out almost the same in the different
 simulations
(Fig. \ref{fig:v2}). These results reflect the universality of the
 transverse flow for different longitudinal pressures \cite{Vredevoogd:2008id}.
Moreover, these observable are determined by the form of the transverse flow at
 the freeze-out. The final transverse flow is built up 
during the whole evolution, not only in the first stage of the collisions, 
when the pressure anisotropy is present.

\begin{figure}[htb]
{\includegraphics[width=0.5\textwidth]{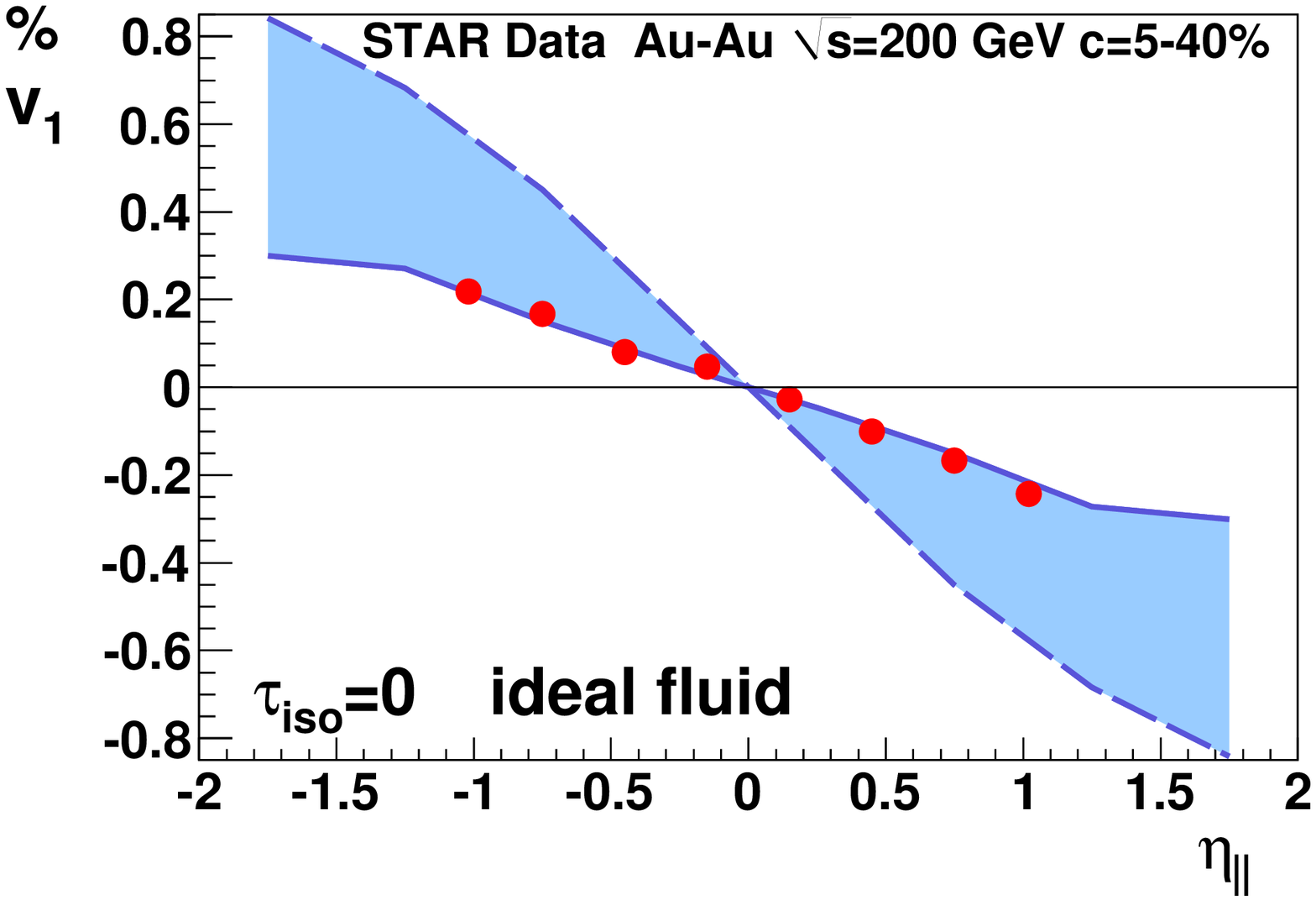}}
{\includegraphics[width=0.5\textwidth]{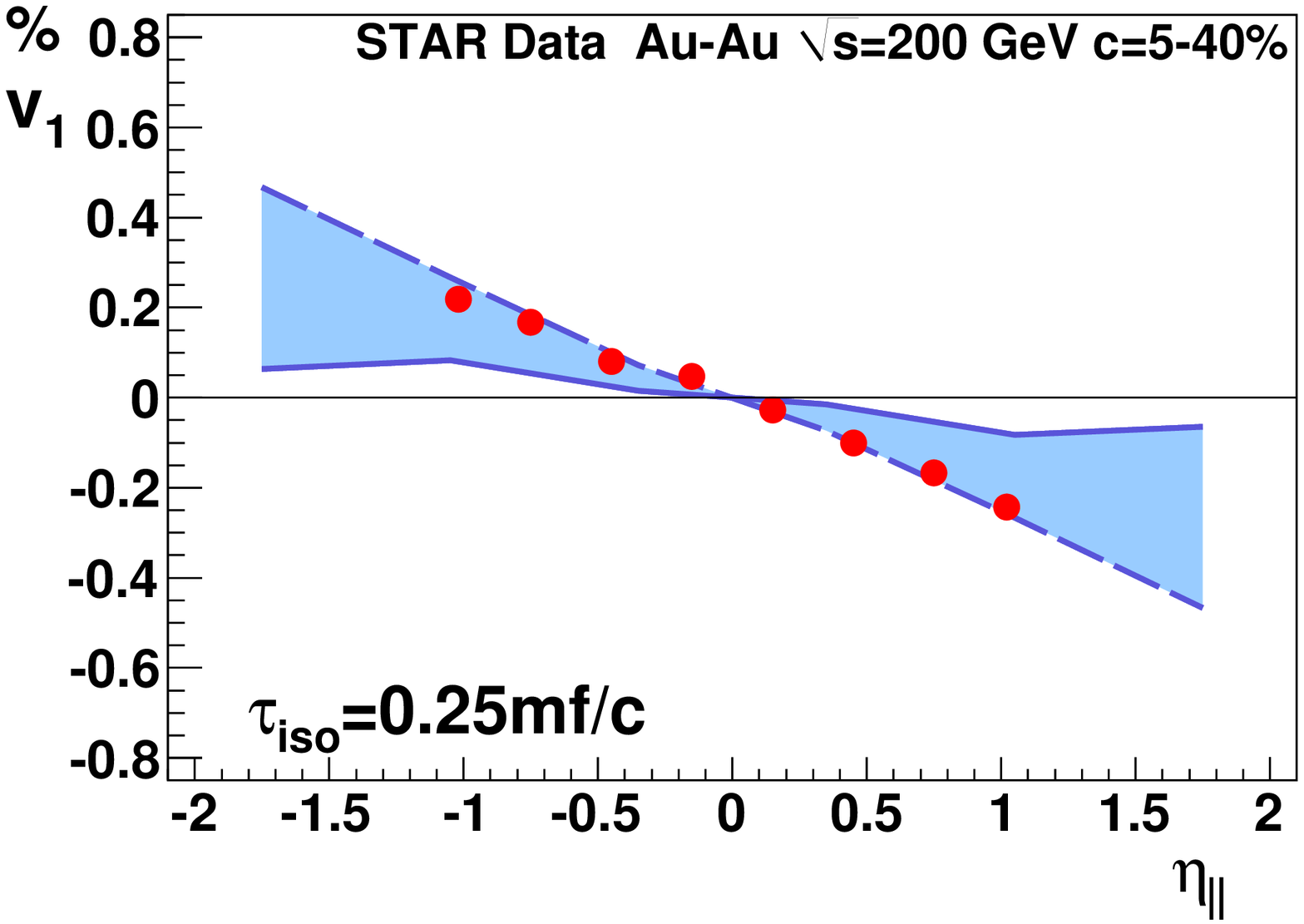}}
\caption{The directed flow calculated in $3+1$ dimensional hydrodynamics with isotropic pressure (left panel) and for the 
pressure anisotropy with a relaxation time $\tau_{iso}=0.25$~fm/c 
(right panel). The shaded band represents the model uncertainty 
related to initial tilt of the source in the Glauber model. 
Data are from  the STAR Collaboration 
 for Au-Au collisions with centrality 
$5$-$40$\% at $\sqrt{s}=200$~GeV (dots) \cite{Abelev:2008jga}. }\label{fig:v1}
\end{figure}

On the other hand, the 
directed flow is a quantity that is very sensitive to the 
longitudinal pressure. The acceleration of the fluid element that
 generates the directed flow requires the simultaneous acceleration 
in the transverse and longitudinal directions
\begin{eqnarray}
{\partial_\tau v_x}=&&-\frac{\partial_x P_\perp}{\epsilon + P} \ , 
\label{eq:ac1} \\
{\partial_\tau Y}=&&-\frac{\partial_{\eta_\parallel} P_\parallel}{\tau(\epsilon + P)} \ ,
\label{eq:ac2} \end{eqnarray}
The direction of the acceleration in the $x$
 direction is usually anti-correlated with the direction 
of the acceleration for the fluid rapidity $Y$. It leads to a negative 
directed flow. Calculations show that the directed flow is formed in 
the first fm/c of the hydrodynamic expansion \cite{Bozek:2010bi}. If in that
 time the longitudinal pressure is reduced, the final directed flow is 
smaller. In Fig. \ref{fig:v1} is shown the calculated directed flow as 
function of pseudorapidity compared to
STAR Collaboration data \cite{Abelev:2008jga}.
In the left panel the results for the perfect fluid hydrodynamics are shown.
 The shaded band represents the uncertainty related to the range of the 
 initial tilt of the fireball. The results for the dynamics with 
asymmetric pressures are plotted in the right panel. Taking into account the uncertainty of the initial conditions we obtain the allowed limits for
the thermalization (isotropization) time 
\begin{equation}
0\le \tau_{iso} \le 0.25\ \mbox{fm/c}\ \ .
\end{equation}

\section{Conclusions}

We study the hydrodynamic expansion of the fireball created 
in a relativistic heavy-ion collision with unequal transverse
 and longitudinal pressures. Using a simple ansatz for the
 pressure anisotropy in the initial stage of the collision,
 we study the possibility of observing the anisotropy.
We show that the transverse momentum spectra, elliptic flow and 
femtoscopy observables are not sensitive to this effect. Instead 
we propose to look at the directed flow generated in the collision.
The directed flow is sensitive to the early pressure anisotropy for two 
reasons,
\begin{itemize}
\item it is formed in the first stage of the expansion,
\item it is built trough a simultaneous action of the transverse 
and longitudinal pressures.
\end{itemize}
The directed flow is reduced if the  longitudinal pressure is smaller. 
Comparison to experimental data sets a limit on the relaxation time for the 
pressure equilibration $\tau_{iso}\le 0.25$~fm/c. The directed flow could 
serve as a sensitive constraint for microscopic models of the 
initial equilibration \cite{Rebhan:2008uj,Bjoraker:2000cf,Xu:2004mz,Chesler:2009cy,Dusling:2010rm,Schenke:2010rr}.

\section*{Acknowledgments}

The work is supported  by the
Polish Ministry of Science and Higher Education 
grant No.  N N202 263438.


\begin{footnotesize}


\begin{thebibliography}{99}


\bibitem{Kolb:2003dz}
P.~F. Kolb and U.~W. Heinz,
 Hydrodynamic description of ultrarelativistic heavy-ion collisions,
 in {\em Quark Gluon Plasma 3}, edited by R.~Hwa and X.~N. Wang, World
  Scientific, Singapore, 2004, nucl-th/0305084.

\bibitem{Hirano:2008aj}
T.~Hirano,
\newblock J. Phys. {\bf G36}, 064031 (2009).

\bibitem{Ollitrault:2010tn}
J.-Y. Ollitrault,
\newblock (2010), arXiv: 1008.3323.

\bibitem{Romatschke:2009im}
P.~Romatschke,
\newblock Int. J. Mod. Phys. {\bf E19}, 1 (2010).

\bibitem{Teaney:2009qa}
D.~A. Teaney,
\newblock (2009), arXiv: 0905.2433.

\bibitem{Broniowski:2008vp}
W.~Broniowski, M.~Chojnacki, W.~Florkowski, and A.~Kisiel,
\newblock Phys. Rev. Lett. {\bf 101}, 022301 (2008).

\bibitem{Pratt:2008qv}
S.~Pratt,
\newblock Phys. Rev. Lett. {\bf 102}, 232301 (2009).

\bibitem{Broniowski:2008qk}
W.~Broniowski, W.~Florkowski, M.~Chojnacki, and A.~Kisiel,
\newblock Phys. Rev. {\bf C80}, 034902 (2009).

\bibitem{Vredevoogd:2008id}
J.~Vredevoogd and S.~Pratt,
\newblock Phys. Rev. {\bf C79}, 044915 (2009).

\bibitem{Bozek:2007di}
P.~Bo\.zek,
\newblock Acta Phys. Polon. {\bf B39}, 1375 (2008).

\bibitem{Florkowski:2010cf}
W.~Florkowski and R.~Ryblewski,
\newblock (2010), arXiv: 1007.0130.

\bibitem{Martinez:2010sc}
M.~Martinez and M.~Strickland,
\newblock Nucl. Phys. {\bf A848}, 183 (2010).

\bibitem{Mauricio:2007vz}
M.~Martinez and M.~Strickland,
\newblock Phys. Rev. Lett. {\bf 100}, 102301 (2008).

\bibitem{Schenke:2006yp}
B.~Schenke and M.~Strickland,
\newblock Phys. Rev. {\bf D76}, 025023 (2007).

\bibitem{Dusling:2008xj}
K.~Dusling and S.~Lin,
\newblock Nucl. Phys. {\bf A809}, 246 (2008).

\bibitem{Bozek:2010aj}
P.~Bo\.zek and I.~Wyskiel-Piekarska,
\newblock (2010), arXiv: 1009.0701.

\bibitem{Bozek:2010bi}
P.~Bo\.zek and I.~Wyskiel,
\newblock Phys. Rev. {\bf C81}, 054902 (2010).

\bibitem{Abelev:2008jga}
STAR, B.~I. Abelev {\em et~al.},
\newblock Phys. Rev. Lett. {\bf 101}, 252301 (2008).

\bibitem{Chojnacki:2007jc}
M.~Chojnacki and W.~Florkowski,
\newblock Acta Phys. Polon. {\bf B38}, 3249 (2007).

\bibitem{Bozek:2009ty}
P.~Bo\.zek and I.~Wyskiel,
\newblock Phys. Rev. {\bf C79}, 044916 (2009).

\bibitem{Bialas:2004su}
A.~Bia\l{}as and W.~Czy\.z,
\newblock Acta Phys. Polon. {\bf B36}, 905 (2005).

\bibitem{Adil:2005qn}
A.~Adil and M.~Gyulassy,
\newblock Phys. Rev. {\bf C72}, 034907 (2005).

\bibitem{Adler:2003cb}
PHENIX, S.~S. Adler {\em et~al.},
\newblock Phys. Rev. {\bf C69}, 034909 (2004).

\bibitem{Adams:2004yc}
STAR, J.~Adams {\em et~al.},
\newblock Phys. Rev. {\bf C71}, 044906 (2005).

\bibitem{Adare:2010ux}
PHENIX, A.~Adare {\em et~al.},
\newblock Phys. Rev. Lett. {\bf 105}, 062301 (2010).

\bibitem{Mrowczynski:2005ki}
S.~Mrowczynski,
\newblock Acta Phys. Polon. {\bf B37}, 427 (2006).

\bibitem{Rebhan:2008uj}
A.~Rebhan, M.~Strickland, and M.~Attems,
\newblock Phys. Rev. {\bf D78}, 045023 (2008).

\bibitem{Bjoraker:2000cf}
J.~Bjoraker and R.~Venugopalan,
\newblock Phys. Rev. {\bf C63}, 024609 (2001).

\bibitem{Xu:2004mz}
Z.~Xu and C.~Greiner,
\newblock Phys. Rev. {\bf C71}, 064901 (2005).

\bibitem{Chesler:2009cy}
P.~M. Chesler and L.~G. Yaffe,
\newblock Phys. Rev. {\bf D82}, 026006 (2010).

\bibitem{Dusling:2010rm}
K.~Dusling, T.~Epelbaum, F.~Gelis, and R.~Venugopalan,
\newblock (2010), arXiv: 1009.4363.

\bibitem{Schenke:2010rr}
B.~Schenke, S.~Jeon, and C.~Gale,
\newblock (2010), arXiv: 1009.3244.


\end{thebibliography}

\end{footnotesize}


\end{document}